\documentclass[runningheads]{llncs}
\usepackage[T1]{fontenc}
\usepackage{graphicx}
\usepackage{hyperref}
\usepackage{tabularx}
\usepackage{booktabs} 
\usepackage{orcidlink}
\begin{document}
\title{Designing and Evaluating Next-Generation Learning Interfaces: Linking AI, HCI, and the Learning Sciences}
\titlerunning{Designing and Evaluating Next-Generation Learning Interfaces}
\newcommand{\IDmengxia}{0000-0002-2676-9032}
\newcommand{\IDyanchen}{0000-0002-1646-6935}
\newcommand{\IDqiaojin}{0000-0001-5493-1343}
\newcommand{\IDyangshi}{0000-0001-6486-4340}
\newcommand{\IDpauldenny}{0000-0002-5150-9806}
\newcommand{\IDtiffanybarnes}{0000-0002-6500-9976}
\newcommand{\IDqingsongwen}{0000-0003-4516-2524}
\newcommand{\IDvincentaleven}{0000-0002-1581-6657}

\author{Meng Xia\inst{1}\orcidlink{\IDmengxia} \and
Yan Chen\inst{2}\orcidlink{\IDyanchen} \and
Qiao Jin\inst{3}\orcidlink{\IDqiaojin}
\and
Yang Shi\inst{4}\orcidlink{\IDyangshi}
\and
Paul Denny\inst{5}\orcidlink{\IDpauldenny}
\and
Tiffany Barnes\inst{3}\orcidlink{\IDtiffanybarnes}
\and
Qingsong Wen\inst{6}\orcidlink{\IDqingsongwen}
\and
Vincent Aleven\inst{7}\orcidlink{\IDvincentaleven}
}
\authorrunning{Xia et al.}
\institute{
Texas A\&M University, TX, USA\and
Virginia Tech, VA, USA\and
North Carolina State University, NC, USA\and
Utah State University, UT, USA\and
University of Auckland, New Zealand\and
University of Oxford, UK\and
Carnegie Mellon University, PA, USA \\
\email{mengxia@tamu.edu}, \email{ych@vt.edu}, \email{qjin4@ncsu.edu}, \email{yang.shi@usu.edu}, \email{paul@cs.auckland.ac.nz}, \email{tmbarnes@ncsu.edu}, \email{qingsongedu@gmail.com}, \email{aleven@cs.cmu.edu}
}
\maketitle              % typeset the header of the contribution
%
% \begin{abstract}
% We propose to host the \textbf{full-day workshop} intended for the \textbf{AIED 2026} conference. 
% Education is being reshaped by rapid advances in generative AI and immersive technologies, which offer unprecedented opportunities for scalable tutoring, personalized feedback, multimodal engagement, and new forms of human–AI collaboration, while also raising challenges such as over-reliance on AI-generated content, erosion of learner agency, issues of trust and bias, and accessibility in immersive environments. Addressing these opportunities and challenges is inherently interdisciplinary, requiring expertise from AI, Human-Computer Interaction (HCI), and learning sciences to ensure that emerging systems are both effective and responsible. This 2026 learning festival workshop brings together researchers from AI, HCI, and learning sciences communities to chart the future of next-generation learning interfaces. Through lightning talks, paper presentations, and collaborative activities, participants will share methods, examine the roles of humans and AI in complex learning scenarios, and identify research agendas that bridge theory and practice, with outcomes including a shared knowledge base of design strategies grounded in learning theory, a roadmap for evaluating emerging educational technologies, and a lasting interdisciplinary community at the intersection of AI, HCI, and education.

% \keywords{Human Computer Interaction \and Educational Technologies \and Interface Design \and Learning Environments}
% \end{abstract}
%
%
%
\section{Type of Event}
This is a full-day, in-person workshop to be held at AIED 2026.

\section{Theme and Goals}
Recent advances in generative AI and immersive technologies, such as augmented and virtual reality, are transforming how people learn, teach, and collaborate across classrooms, homes, and workplaces. These technologies enable on-demand tutoring, adaptive content generation, multimodal interaction, and new forms of human–AI collaboration in shared environments. At the same time, their integration into education introduces critical challenges, including balancing efficiency with learner agency, ensuring accessibility in immersive settings, and addressing concerns related to bias, privacy, and over-reliance on AI. As a result, interaction design plays a central role in determining whether these technologies effectively support learning.

Prior work in intelligent tutoring systems has demonstrated the value of theoretically grounded educational technologies, with successful applications in domains such as mathematics and computer science~\cite{aleven2006cognitive,price2017isnap}. These systems integrate cognitive theory~\cite{aleven2006cognitive}, evidence-based practices~\cite{koedinger2002toward}, and data-driven approaches~\cite{barnes2005q,koedinger2013using,price2017isnap} to achieve measurable learning gains. However, they are typically built on fixed interaction paradigms, limiting their ability to support the open-ended, multimodal, and collaborative learning scenarios enabled by recent advances in AI and immersive technologies.

At the same time, research across AI, HCI, and the learning sciences has made significant progress in isolation. HCI contributes human-centered design principles and novel interaction techniques for educational interfaces~\cite{tang2024vizgroup,pan2025tutorup,zhang2026classaid}. AI provides methods for adaptive modeling, natural language interaction, and generative content creation~\cite{koedinger2013datadriven,piech2015deep,chu2025llm}. Learning sciences offer theoretical and pedagogical foundations that ground technology design in evidence-based practice~\cite{aleven2002metacognitive}. However, these perspectives are often not well integrated in the design and evaluation of next-generation learning interfaces.

This workshop addresses this gap by bringing together researchers and practitioners from AI, HCI, and the learning sciences to explore how interactive systems can better support learning. We focus on the design and evaluation of human–AI collaborative learning interfaces that are technically robust, human-centered, and pedagogically grounded. By fostering interdisciplinary dialogue, the workshop aims to identify shared challenges, design principles, and research directions for next-generation learning technologies.

The workshop has three specific aims, each accompanied by guiding questions to scaffold discussion and collaboration:  

\begin{itemize}
    \item \textbf{Engage AI and HCI researchers in educational design challenges.}  
    We aim to encourage the application of novel interactive technologies (e.g., intelligent interfaces, immersive environments, and generative AI systems) to real learning contexts.  
    % \begin{itemize}
    %     \item How can emerging interaction paradigms (e.g., multimodal input, embodied interaction) be adapted to support learning across diverse contexts?  
    %     \item What role should generative AI and immersive technologies play in complementing, rather than replacing, human teachers and peers?  
    %     \item How do we design for accessibility while introducing advanced technologies in classrooms, workplaces, or informal learning settings?  
    % \end{itemize}

    \item \textbf{Introduce learning science theories and methods.}  
    We aim to surface theories and approaches (e.g., student modeling, learning analytics, educational data mining) that can inform more data-driven and theoretically grounded interface and interaction design.  
    % \begin{itemize}
    %     \item What learning theories (e.g., constructivism, cognitive apprenticeship) can most effectively guide design with generative AI and AR/VR?  
    %     \item How should we model learner knowledge, skills, and affect in real time, and how might these models adapt across modalities?  
    %     \item What ethical, privacy, and trust concerns emerge when collecting and analyzing large-scale learner data in immersive or AI-enhanced environments?  
    % \end{itemize}

    \item \textbf{Bridge AI, HCI, and learning sciences communities.}  
    We seek to inspire new collaborations and cross-community research agendas that advance both the design and the impact of educational technologies.  
    % \begin{itemize}
    %     \item How can we create sustainable infrastructures and shared datasets that serve both HCI, AI, and learning sciences communities?  
    %     \item What practices can help us balance methodological rigor from learning sciences with the rapid prototyping culture of HCI?  
    %     \item How can collaborations between researchers, educators, and technologists be structured to ensure mutual benefit and long-term impact?  
    % \end{itemize}
\end{itemize}

% \section{Workshop Plans and Activities}
% This \textbf{in-person full-day} workshop will bring together approximately 30 participants (including organizers). Participants will present their position/work-in-progress papers, engage with experienced mentors from AI, HCI, and learning sciences, and collaboratively explore challenges and opportunities in interdisciplinary educational technology research.

\begin{table}[t]
  \caption{Workshop Day Schedule}
  \label{tab:workshop-schedule}
  \small
  \centering
  \begin{tabularx}{\columnwidth}{@{}l l X@{}}
    \toprule
    \textbf{Session} & \textbf{Time} & \textbf{Activity / Description} \\
    \midrule
    Introduction & 09:00--10:00 & Workshop introduction and participants introduction, including their background, current work, and potential contribution (e.g., dataset, tool, design method). \\
    Keynote & 10:00--10:45 & Invited keynote aligned with the workshop theme. \\
    Spotlight Presentations I & 10:45--12:00 & Selected submissions (short talks): 8 min talk + 4 min feedback; quick survey after each to surface connections (methods, collaborators, evaluation ideas). \\
    \textit{Lunch Break} & 12:00--14:00 & Organizer-hosted group lunch near the venue. \\
    Spotlight Presentations II & 14:00--14:45 & Continuation of selected submissions with the same format (about 10 total across both sessions). \\
    Thematic Breakouts & 14:45--15:45 & Curated themes with facilitation and note-taking: (1) controllability of educational interfaces; (2) scalable feedback; (3) data \& privacy; (4) evaluation paradigms; plenary share-out. \\
    Speed Collaboration Rounds & 16:00--16:45 & Paired 10-minute exchanges across disciplines; rotate to surface concrete collaboration opportunities. \\
    Closing Reflections & 16:45--17:00 & Synthesis of insights and next steps for continued collaboration. \\
    % \textit{Networking Dinner (Optional)} & 18:00--19:00 & Informal dinner near the venue. \\
    \bottomrule
  \end{tabularx}
\end{table}

\section{Workshop Planned Activities}
% See Table~\ref{tab:activities}
%Consider the community (send out survey)
 % List of conferences attended
 % What types of input you want to get from the other community?
%Collect feedback on matching/topics to discuss
A tentative schedule is shown in Table 1.
\subsubsection{Introduction (9:00 - 10:00)}
The workshop will begin with an introduction, followed by participants' introductions. The host will play and introduce the slide deck about the participants' information.

\subsubsection{Keynote (10:00 - 10:45):} We will invite a keynote speaker to share insights related to the workshop theme. 
% Potential keynote invitees include Juho Kim, whose work at KAIST bridges HCI and education through learner-centered interaction design and scalable platforms; Alice Oh, an expert in human-centered AI and data-driven methods with implications for responsible educational technologies; Peter Brusilovsky, a pioneer of adaptive hypermedia and personalized learning with decades of contributions to intelligent tutoring systems; and Chris Piech, a leader in computing education at Stanford, known for AI-driven personalized learning and programming education research. Together, they represent diverse but complementary perspectives across HCI, AI, and the learning sciences, making them ideal to open the workshop conversation

\subsubsection{Paper Presentation (10:45 - 12:00, 2:00 - 2:45)}
Following the keynote, 10 selected workshop submissions will be invited to present in a short spotlight format in two sections. Each presenter will have 8 minutes to share their work, followed by 4 minutes of feedback from the audience. 
% The presentations will highlight projects at the intersection of HCI, AI, and learning sciences. 
A quick feedback survey will be given after each talk, encouraging attendees to identify interdisciplinary connections and complementary methods that could improve the work, such as: potential collaborators, methods for improving data quality, expanded interface affordances or stakeholder considerations, or broader evaluation strategies. 
% This activity aims to promote cross-pollination between communities and drive stronger and more holistic research.

\subsubsection{Thematic Breakout Discussions: ``Where Worlds Collide'' (2:45 - 3:45)}
In the afternoon, we will organize a breakout session where participants will form small groups to discuss challenges and opportunities at the intersection of educational technology and user interface research. Each group will focus on one of several curated themes, including: (1) Controllability in educational interfaces - How we can enable educators or students to guide or refine AI behavior; (2) Scalable feedback mechanisms - How interface design can support timely and personalized feedback in large-scale learning environments; (3) Data and privacy - Balance the need for detailed educational data with transparency, consent, and trust; and (4) Evaluation paradigms - Understanding how HCI and EdTech communities approach evaluation differently, and what we can learn from each other. Each breakout group will be facilitated and will record key ideas to share in a whole-group reconvening session.

\subsubsection{Speed Collaboration Rounds (4:00-4:45)}
To foster cross-community connections, we will host a speed collaboration activity. Participants from different disciplinary backgrounds will be paired for short, timed conversations. Each person will introduce their work and describe a specific problem they are trying to solve, followed by a short discussion of how their partner might contribute. For example, a participant working on educational data mining may offer a dataset and seek help designing an interface to visualize model output, while a HCI researcher may describe a novel interaction technique and ask for feedback on how it could be deployed in a classroom context. After each 10-minute round, the participants rotate to meet someone new. 
% This activity is intended to surface opportunities for mutual exchange and spark new collaborations.

\subsubsection{Closing Reflections (4:45-5:00).} We will conclude the workshop with a brief reflection session, where the organizers will summarize insights from the day and outline the next steps to continue the conversation beyond the workshop.

% \subsection{Post-Workshop Plans}
% % Our anticipated outcomes include:
% \paragraph{Workshop Report}
% We will create a concise report capturing key ideas, insights, and themes from presentations and group discussions. This report will include participant contributions and will be openly accessible online, providing a valuable resource for both attendees and the broader research community. Participants will also have the option to publish their position papers alongside the report on platforms such as arXiv.

% % \paragraph{Interdisciplinary Knowledge Base and Toolkit}
% % A primary outcome will be a collaboratively-developed online knowledge base synthesizing learning theories, and educational data-driven methods from learning science community. By analyzing participants' position papers and interactive workshop discussions, we will extract actionable insights and best practices. This resource will be published on the workshop website.

% \paragraph{Building a Lasting Community}
% We will maintain the communication channels (Slack and Google Groups) to enable collaboration beyond the event itself. In addition, participants will collaboratively review the workshop report, sustaining engagement and supporting lasting professional connections. We will also organize follow-up workshops at other AI, HCI, and learning sciences conferences to build the community across disciplines.

\section{Prior or Related Work}
This is the first edition of this workshop.

Recent years have seen increasing attention to AI-supported learning across multiple research communities. Major venues such as NeurIPS, AAAI, and KDD have hosted workshops on AI for Education\footnote{\url{https://gaied.org/neurips2023/index.html}}\footnote{\url{https://ai4ed.cc/workshops/aaai2022}}\footnote{\url{https://ai-for-edu.github.io/workshop_kdd2024.html}}
, while CHI has explored topics such as augmented educators and the future of work\footnote{\url{https://sites.google.com/view/augemted-educators-and-ai/home}}
. Learning sciences and learning analytics communities, including ICLS, CSCL, and LAK, have also advanced work on pedagogy and data-driven learning, with recent workshops on large language models for qualitative research and generative AI in learning analytics\footnote{\url{https://sites.google.com/view/lak-25-workshop-llms-for-qual/}}\footnote{\url{https://sites.google.com/monash.edu/genai-la-workshop-lak25/}}. These efforts highlight growing interest in leveraging AI technologies to support learning. However, they are often fragmented across communities. AI research tends to focus on modeling and system capabilities, HCI emphasizes interaction design and user experience, and the learning sciences prioritize pedagogical theory and learning outcomes. As a result, there remains a lack of venues that explicitly integrate these perspectives to guide the design and evaluation of next-generation learning interfaces. This workshop builds on these prior efforts while focusing on bridging AI, HCI, and the learning sciences. It aims to provide a dedicated space for interdisciplinary dialogue and collaboration around human–AI collaborative learning systems.

\section{Program Committee}
Our organizing committee brings together an interdisciplinary group of scholars spanning HCI, AI, and the learning sciences.
% Collectively, the team has deep expertise in human–AI interaction, data visualization, programming support tools, real-time learning analytics, XR and mixed-reality learning environments, intelligent tutoring systems, collaborative and student-generated learning, computing education research, educational data mining, learning games, and large-scale educational platforms. This breadth positions the workshop to foster meaningful cross-disciplinary dialogue and advance both theory and practice.

\textbf{Meng Xia} is an Assistant Professor in Computer Science and Engineering at Texas A\&M University. Her research interests include Human–AI Interaction, Data Visualization, and Educational Technology, with a focus on human–AI collaboration for personalized education. She will serve as the General Chair and will coordinate all aspects of the workshop.

\textbf{Yan Chen} is an Assistant Professor of Computer Science at Virginia Tech. His work spans programming support tools, real-time learning analytics, and learning at scale, with a focus on interactive Human–AI systems for education. He will chair website development and technical infrastructure.

\textbf{Qiao (Georgie) Jin} is an Assistant Professor of Computer Science at North Carolina State University. Her research explores XR- and AI-driven mixed-reality tools to support teaching, learning, and social connection, particularly in real-world educational settings.

\textbf{Yang Shi} is an Assistant Professor of Computer Science at Utah State University. His research focuses on data-driven representations of program code for intelligent tutoring systems and student modeling in computing education, drawing on data mining and machine learning approaches.

\textbf{Paul Denny} is an ACM Distinguished Member and Professor at the University of Auckland whose work centers on collaborative learning and student-generated content in computing education. He is the creator of PeerWise, a large-scale platform used internationally, and brings extensive experience in building sustained research communities.

\textbf{Tiffany Barnes} is a Distinguished Professor of Computer Science at North Carolina State University. Her research focuses on computing education, educational data mining, and AI-supported learning environments. She will lead workshop outreach and community engagement, drawing on her extensive experience in inclusive computing initiatives.

\textbf{Qingsong Wen} is Head of AI and Chief Scientist at Squirrel Ai Learning and a PhD Supervisor at the University of Oxford. His research spans machine learning, time-series analysis, and AI for education, with extensive leadership experience across major AI conferences and professional societies.

\textbf{Vincent Aleven} is a Professor at Carnegie Mellon University’s Human-Computer Interaction Institute and Director of the CATS Lab. His work focuses on intelligent tutoring systems, learning analytics, and authoring tools for educational technologies. He will co-lead workshop dissemination and outreach.

\bibliographystyle{splncs04}
\bibliography{sample-base}

@inproceedings{aleven2006cognitive,
  title={The cognitive tutor authoring tools (CTAT): Preliminary evaluation of efficiency gains},
  author={Aleven, Vincent and McLaren, Bruce M and Sewall, Jonathan and Koedinger, Kenneth R},
  booktitle={Intelligent Tutoring Systems: 8th International Conference, ITS 2006, Jhongli, Taiwan, June 26-30, 2006. Proceedings 8},
  pages={61--70},
  year={2006},
  organization={Springer}
}

@inproceedings{price2017isnap,
  title={iSnap: towards intelligent tutoring in novice programming environments},
  author={Price, Thomas W and Dong, Yihuan and Lipovac, Dragan},
  booktitle={Proceedings of the 2017 ACM SIGCSE Technical Symposium on computer science education},
  pages={483--488},
  year={2017}
}

@inproceedings{chu2025llm,
  title={Llm agents for education: Advances and applications},
  author={Chu, Zhendong and Wang, Shen and Xie, Jian and Zhu, Tinghui and Yan, Yibo and Ye, Jinheng and Zhong, Aoxiao and Hu, Xuming and Liang, Jing and Yu, Philip S and Wen, Qingsong},
  booktitle={EMNLP},
  year={2025}
}

@inproceedings{koedinger2013using,
  title={Using data-driven discovery of better student models to improve student learning},
  author={Koedinger, Kenneth R and Stamper, John C and McLaughlin, Elizabeth A and Nixon, Tristan},
  booktitle={Artificial Intelligence in Education: 16th International Conference, AIED 2013, Memphis, TN, USA, July 9-13, 2013. Proceedings 16},
  pages={421--430},
  year={2013},
  organization={Springer}
}

@inproceedings{koedinger2002toward,
  title={Toward Evidence for Instructional Design Principles: Examples from Cognitive Tutor Math 6},
  author={Koedinger, Kenneth R},
  booktitle={Annual Meeting [of the] North American Chapter of the International Group for the Psychology of Mathematics Education},
  year={2002}
}

@inproceedings{barnes2005q,
  title={The Q-matrix method: Mining student response data for knowledge},
  author={Barnes, Tiffany},
  booktitle={American association for artificial intelligence 2005 educational data mining workshop},
  pages={1--8},
  year={2005},
  organization={AAAI Press, Pittsburgh, PA, USA}
}

@inproceedings{tang2024vizgroup,
  title={VizGroup: An AI-assisted Event-driven System for Collaborative Programming Learning Analytics},
  author={Tang, Xiaohang and Wong, Sam and Pu, Kevin and Chen, Xi and Yang, Yalong and Chen, Yan},
  booktitle={Proceedings of the 37th Annual ACM Symposium on User Interface Software and Technology},
  pages={1--22},
  year={2024}
}

@inproceedings{pan2025tutorup,
  title={Tutorup: What if your students were simulated? training tutors to address engagement challenges in online learning},
  author={Pan, Sitong and Schmucker, Robin and Garcia Bulle Bueno, Bernardo and Llanes, Salome Aguilar and Albo Alarc{\'o}n, Fernanda and Zhu, Hangxiao and Teo, Adam and Xia, Meng},
  booktitle={Proceedings of the 2025 CHI Conference on Human Factors in Computing Systems},
  pages={1--18},
  year={2025}
}

@article{aleven2002metacognitive,
  title={An effective metacognitive strategy: Learning by doing and explaining with a computer-based Cognitive Tutor},
  author={Aleven, Vincent and Koedinger, Kenneth R},
  journal={Cognitive Science},
  volume={26},
  number={2},
  pages={147--179},
  year={2002},
  publisher={Wiley Online Library}
}

@inproceedings{koedinger2013datadriven,
  title={Using data-driven discovery of better student models to improve student learning},
  author={Koedinger, Kenneth R and Stamper, John C and McLaughlin, Elizabeth A and Nixon, Tristan},
  booktitle={Artificial Intelligence in Education: 16th International Conference, AIED 2013, Memphis, TN, USA, July 9-13, 2013. Proceedings},
  volume={7926},
  pages={421--430},
  year={2013},
  organization={Springer}
}

@inproceedings{piech2015deep,
  title={Deep knowledge tracing},
  author={Piech, Chris and Bassen, Joel and Huang, Jonathan and Ganguli, Surya and Sahami, Mehran and Guibas, Leonidas J and Sohl-Dickstein, Jascha},
  booktitle={Advances in Neural Information Processing Systems},
  volume={28},
  year={2015}
}

@article{zhang2026classaid,
  title={ClassAid: A Real-time Instructor-AI-Student Orchestration System for Classroom Programming Activities},
  author={Zhang, Gefei and Sun, Guodao and Xia, Meng and Liang, Ronghua},
  journal={arXiv preprint arXiv:2602.06734},
  year={2026}
}
\end{document}